\documentclass[electronic]{vgtc}             

\graphicspath{{figures/}{pictures/}{images/}{./}} 

\usepackage{times}                     

\usepackage{tabu}                      
\usepackage{booktabs}                  
\usepackage{lipsum}                    
\usepackage{mwe}                       

\usepackage{graphicx}
\usepackage[utf8]{inputenc}
\DeclareUnicodeCharacter{02BB}
{\raisebox{1.5ex}{\rotatebox[origin=c]{180}{,}}}

\usepackage{mathptmx}                  

\onlineid{0}

\vgtccategory{Application or Design Study}

\vgtcinsertpkg

\title{Illustrating Transition Scenarios to Renewable Energy in Hawaiʻi with ProjecTable}

\author{Tracy Bui\thanks{e-mail: buit@hawaii.edu}\\ %
        \scriptsize University of Hawaiʻi at Mānoa\\
        \scriptsize Laboratory for Advanced Visualization \& Applications
\and Kari Noe\thanks{e-mail: karinoe@hawaii.edu}\\ %
        \scriptsize University of Hawaiʻi at Mānoa\\
        \scriptsize Laboratory for Advanced Visualization \& Applications\\
        \scriptsize Create(x)
\and Marissa Halim \thanks{e-mail: mhalim@hawaii.edu}\\ %
        \scriptsize University of Hawaiʻi at Mānoa\\
        \scriptsize Laboratory for Advanced Visualization \& Applications\\
        \scriptsize Creativity and Technology Lab
\and Nurit Kirshenbaum\thanks{e-mail: nuritk@hawaii.edu}\\ %
        \scriptsize University of Hawaiʻi at Mānoa\\
        \scriptsize Laboratory for Advanced Visualization \& Applications\\
        \scriptsize Creativity and Technology Lab
\and Jason Leigh\thanks{e-mail: leighj@hawaii.edu}\\ %
        \scriptsize University of Hawaiʻi at Mānoa\\
        \scriptsize Laboratory for Advanced Visualization \& Applications\\
        \scriptsize Create(x)
}

\teaser{
  \centering
  \includegraphics[width=0.32\linewidth]{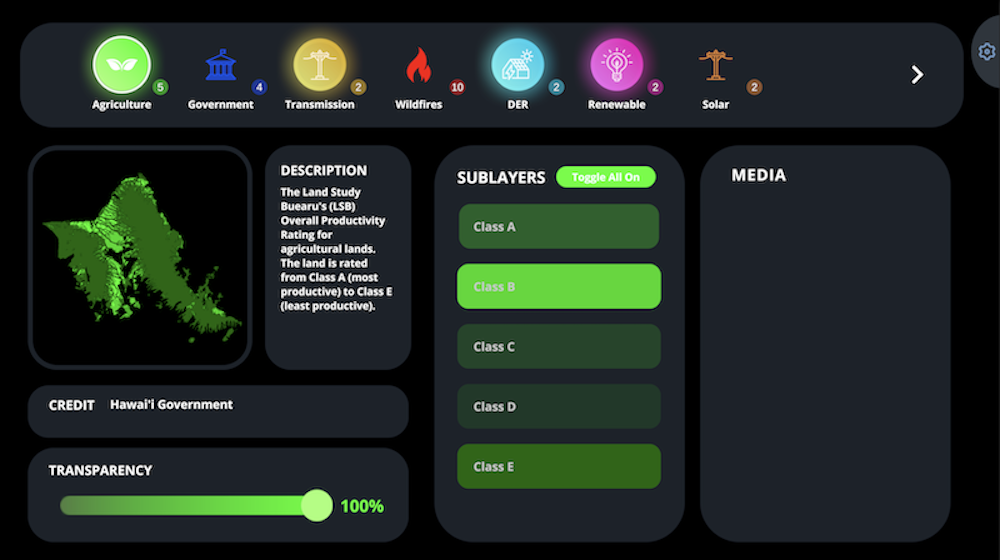}
  \includegraphics[width=0.32\linewidth]{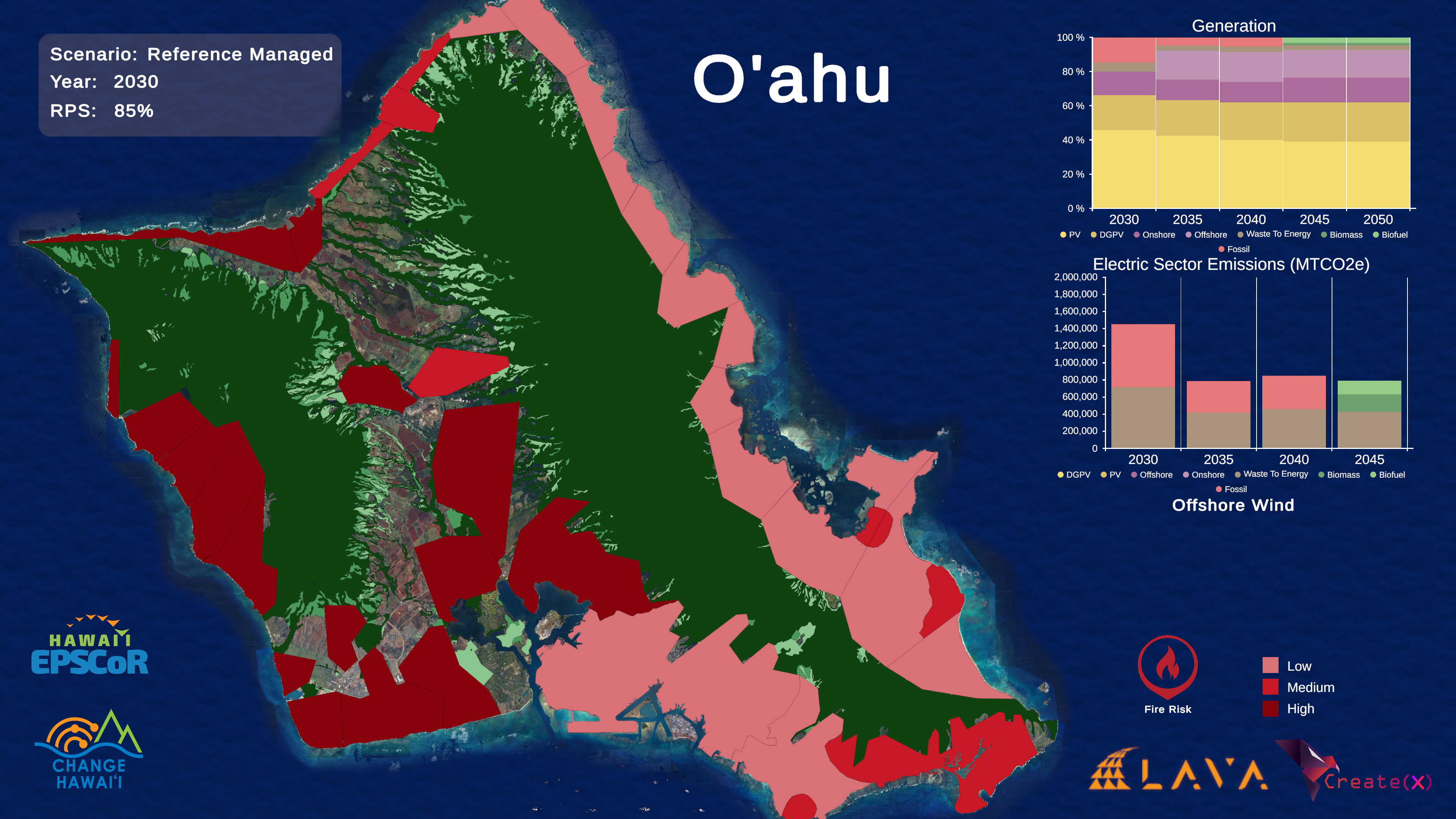}
  \includegraphics[width=0.32\linewidth]{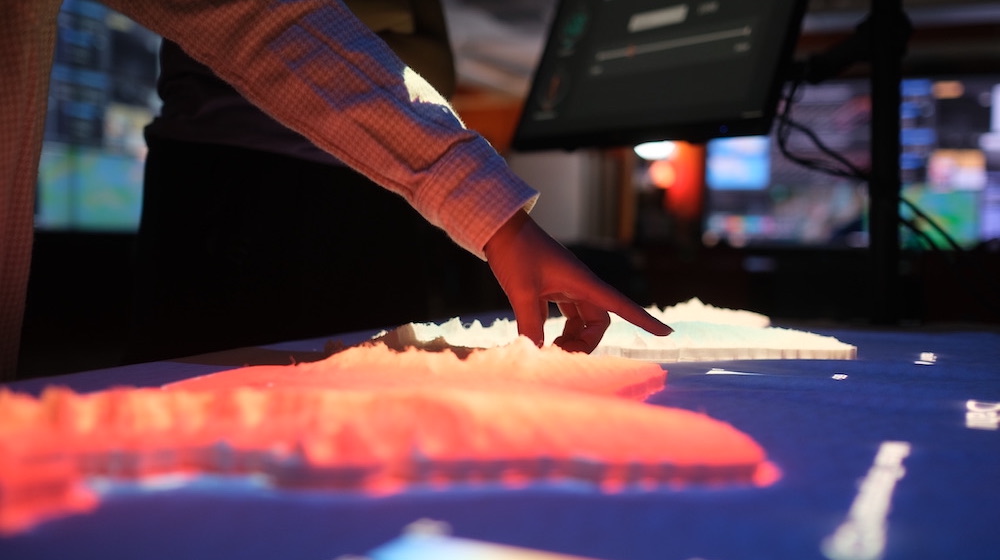}
  \caption{(Left) Touch screen display of the ProjecTable 3.0 system showing the agriculture data layer, icon highlight indicates both this layer, Transmission, DER, and Renewable layers are active on the table display. The sublayer buttons can be used to activate and deactivate specific sublayers, a slider can adjust the visibility of the layer. (Center) Screen shot of the view from the Table display, showing the agricultural and fire risks layers on top of the satellite base map, additional charts are shown at the top-right section of the table. (Right) A view of the Table display projected on top of a 3D printed model of the Oʻahu's terrain showing rainfall data.}
  \label{fig:teaser}
}

\abstract{
     Creating engaging and immersive data visualization tools has become increasingly significant for a wide range of users who want to display their data in a meaningful way. However, this can be limiting for individuals with varying levels of coding expertise. There are specific needs, such as visualizing complex data in easily understandable ways, highlighting real-world problems, or telling a story with data. The Makawalu Visualization Environment (VE) package aims to address these needs through three distinct modular tools: Author, Presenter, and Editor. These tools work together to facilitate different use cases based on the user's requirements. This paper discusses the latest version of the ProjecTable and focuses on the design and usage of the Makawalu VE Author and Presenter tools.
} 

\keywords{Geospatial Visualization, Physicalization, Interactive Tools, Community Co-Design, Modular UI Components.}

\begin{document}


\firstsection{Introduction}

\maketitle

The Makawalu VE tools originate from the Laboratory for Advanced Visualization and Applications (LAVA)’s ProjecTable system \cite{10.1145/3334480.3382968}. This system is a data visualization and physicalization tool that has undergone multiple design iterations. The current version integrates a touchscreen display, allowing users to view and toggle the visibility of multiple geospatial data layers. It is connected to a projector that overlays data onto a 3D model placed on the table.

Since its creation, the ProjecTable has been adapted for specific use cases, including two iterations developed for the Hawai'i State Energy Office (HSEO) and the Hawaiian Electric Company (HECO). These iterations were designed to communicate with the general public and policymakers about pathways to achieving Hawai'i’s goal of 100\% renewable energy by 2045 \cite{HSEO23}. Using land-use data and scenarios, these versions enabled users to visualize trade-offs and resource availability, fostering informed discussions about sustainable energy strategies.

One of the ProjecTable's strengths is its ability to present complex data in an intuitive and immersive format. By projecting 2D geospatial data onto a physical 3D model, users could better understand the information and feel a deeper connection to the data. This innovative approach enhanced engagement and accessibility, making it a powerful tool for fostering dialogue and understanding among diverse audiences.

Ongoing collaboration with HSEO has allowed LAVA to continuously update and refine HSEO's in-house ProjecTable system. Research assistants at LAVA typically hardcoded requested features, such as adding new data layers or charts. While this process enabled customization, it required significant time and developer resources to process files and integrate new information.

\begin{figure}[tb]
 \centering 
    \includegraphics[width=\columnwidth]{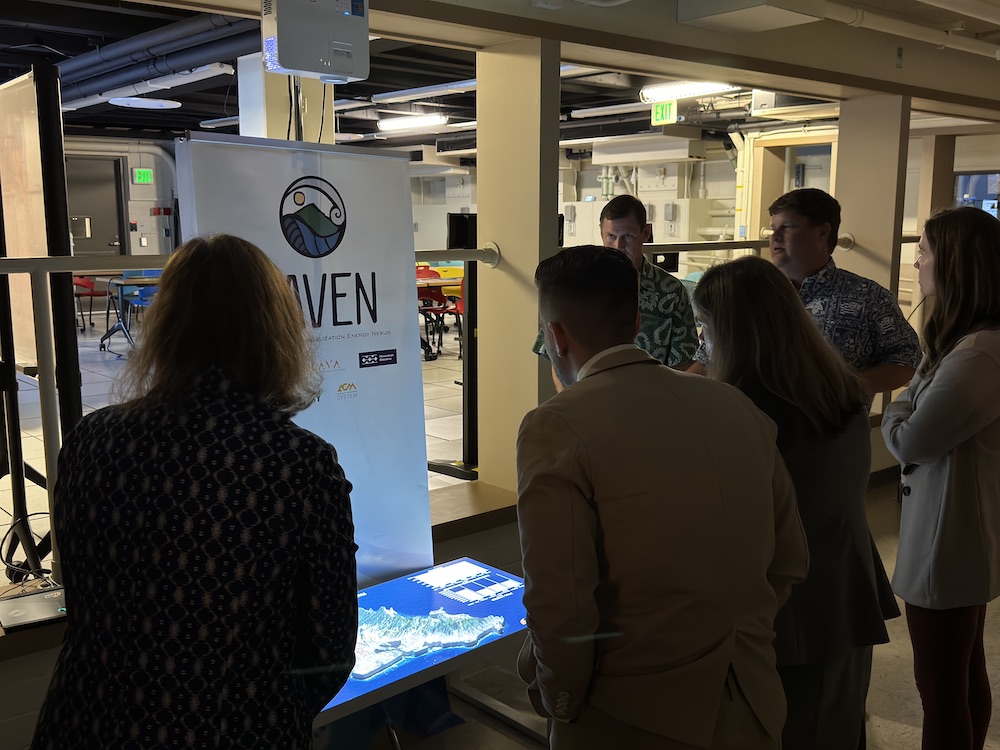}
    \includegraphics[width=\columnwidth]{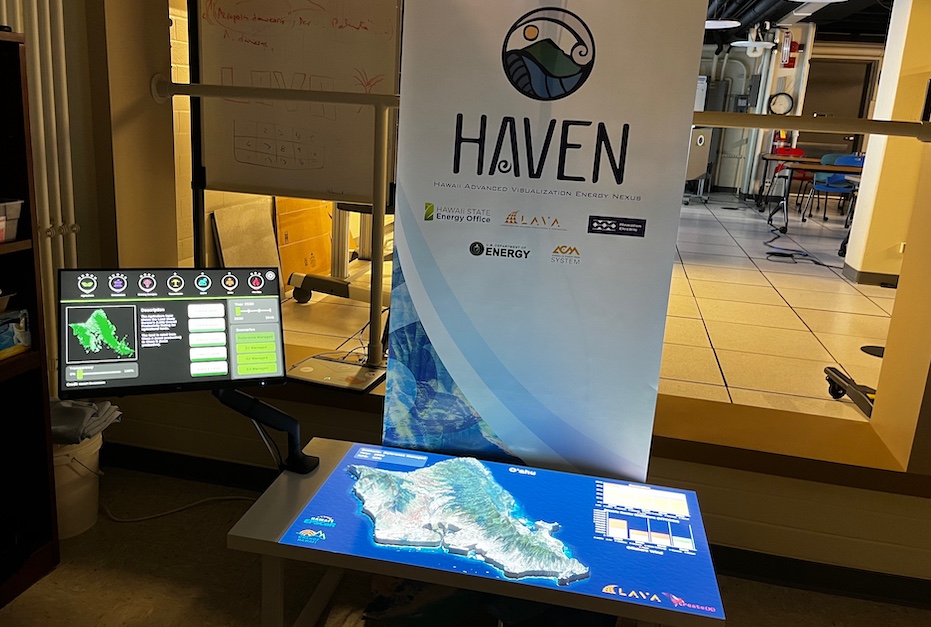}
 \caption{ Photos from the NASEO tour at LAVA: (Top) The HAVENTable being used by HSEO representatives to explain various scenarios for achieving sustainability in Hawai'i. (Bottom) The HAVENTable system during the tour, featuring a lower table setup to explore alternative table positions for optimal projection. }
 \label{fig:naseoTour}
\end{figure}

\begin{figure}[tb]
 \centering 
    \includegraphics[width=\columnwidth]{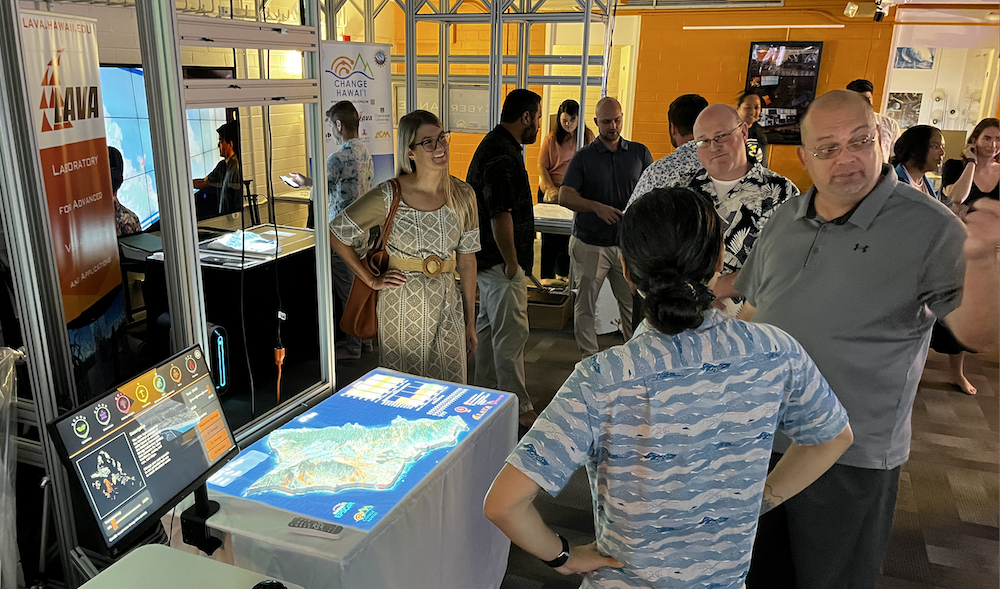}
    \includegraphics[width=\columnwidth]{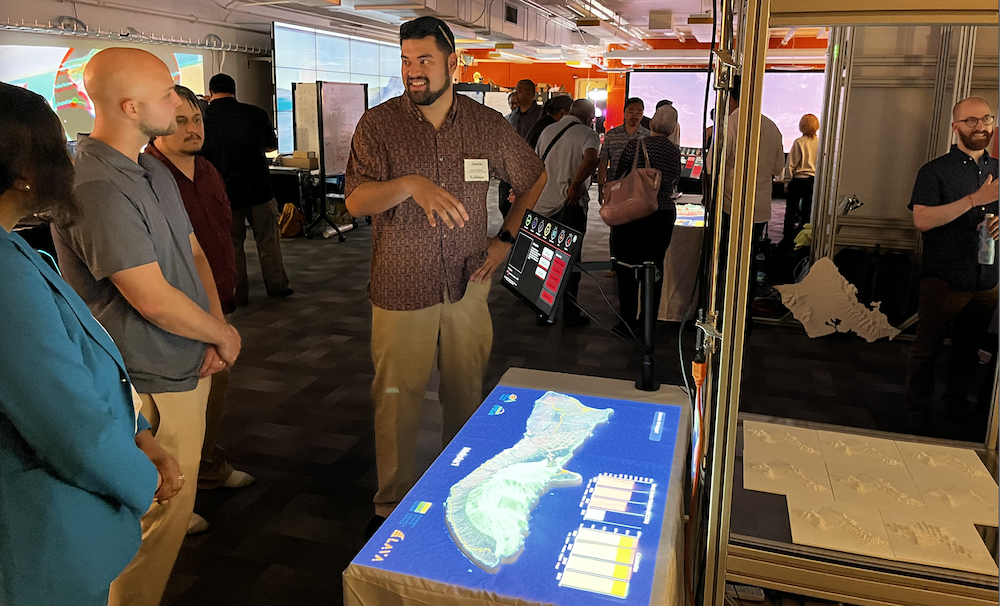}
 \caption{ Photos from the NGA tour at LAVA: (Top) HSEO representative discussing O`ahu-related data with visitors. (Bottom) HSEO representative presenting data related to Moloka`i to the visitors. }
 \label{fig:ngaTour}
\end{figure}

As the ProjecTable became more portable and was showcased at schools and conferences, its popularity grew. It has been toured and taken to numerous venues, including the National Association of State Energy Officials (NASEO) in 2024 and previous years (see Figure~\ref{fig:naseoTour}), National Governors Association (NGA) (See Figure~\ref{fig:ngaTour}), the 13th Festival of Pacific Arts and Culture (FestPAC) 2024, the Catalyst for Change: Indigenous Innovation Conference 2024, and the American Geophysical Union (AGU) 2023. However, the increasing number of users with unique data requirements highlighted the limitations of the hardcoding approach. Updating and customizing multiple ProjecTable applications for diverse audiences proved impractical and resource-intensive. To address these challenges, LAVA developed the third iteration of the ProjecTable, which leverages the Makawalu Visualization Environment (VE). This modular system consists of three separate tools: Author, Presenter, and Editor. The Makawalu VE tools empower users to create, share, and customize their own ProjecTable versions with greater ease and flexibility, meeting the growing demand for a scalable and user-friendly solution.

\section{ProjecTable System}
This work will briefly describe the previous iterations the ProjecTable systems and then go over the third iteration of ProjecTable and its current design and interface. The ProjecTable is the hardware architecture that the Makawalu VE tools and was designed for. 

\subsection{Background} 
The ProjecTable system is drawn from the idea of Data Physicalization \cite{10.1145/2702123.2702180} and Spatial Augmented Reality (SAR) \cite{bimber2005spatial}. The system that is dicussed throughout this paper is a product of many iterative design cycles that started with a proof-of-concept protoptype where researchers at LAVA manually held a small pico projector above a small 3D printed model of O`ahu. This led to creating custom hardware setup with a larger 3D printed model with tangible pucks to control the user interface, also known as the ProjecTable 1.0 system \cite{10.1145/3334480.3382968, 10.1145/3311790.3396630}. 

\subsection{Prior Iterations}
\subsubsection{ProjecTable 1.0}
ProjecTable 1.0, also referred to as the ``HAVENTable,’’ originated from the HAVEN (Hawai‘i Advanced Visualization Energy Nexus) project developed at the Laboratory for Advanced Visualization and Applications (LAVA) at UH Mānoa, led by students James Hutchison III and Ryan Theriot. This system was created in collaboration with HSEO and HECO to support Hawai‘i's renewable energy goals, specifically targeting 100\% sustainability by 2045. The primary aim of ProjecTable 1.0 was to help HSEO communicate complex land-use and energy transition scenarios to the public. The ProjecTable 1.0 system has been shown at the NASEO 2019 Annual Meeting, showcasing a new development of data visualization and physicalization. 

\begin{figure}[tb]
 \centering 
    \includegraphics[width=\columnwidth]{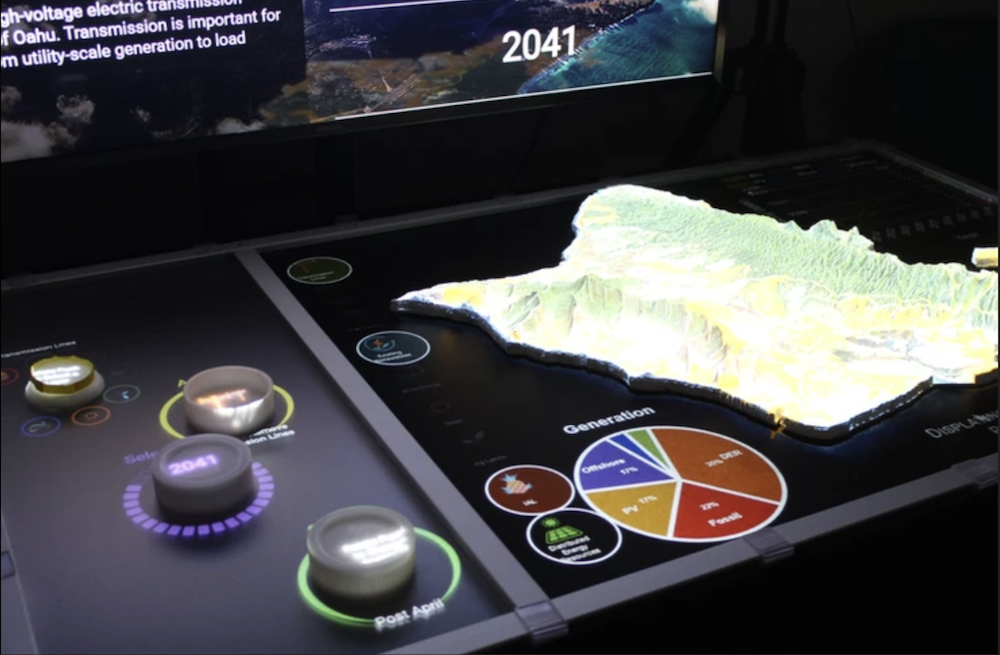}
 \caption{ The ProjecTable 1.0 system with tangible pucks currently dialed to the year 2041 and displaying data related to the Post April scenario. }
 \label{fig:projectable1}
\end{figure}

The system utilized tangible, 3D-printed pucks to interact with the data. These pucks served as intuitive controls for navigating geospatial data layers, exploring various scenarios, and visualizing changes over time. Movement of the pucks was tracked and translated into visual and informational outputs on the table. The structure of ProjecTable 1.0 featured a robust design with large poles supporting an overhead projector. The focus of this iteration was on showcasing initial renewable energy scenarios developed under the HAVEN project. See Figure~\ref{fig:projectable1}.

\subsubsection{ProjecTable 2.0}
ProjecTable 2.0 represented a significant evolution from ProjecTable 1.0, incorporating lessons learned from HAVEN and expanding upon the initial concept. This iteration was funded by HSEO, and the Department of Energy (DOE), and commissioned by HECO for their Illumination event in 2022. The system has demonstrated broader organizational support and interest in applying the system to new challenges. With a table twice the size of the original, ProjecTable 2.0 was capable of visualizing more complex and diverse data sets.

\begin{figure}[tb]
 \centering
    \includegraphics[width=\columnwidth]{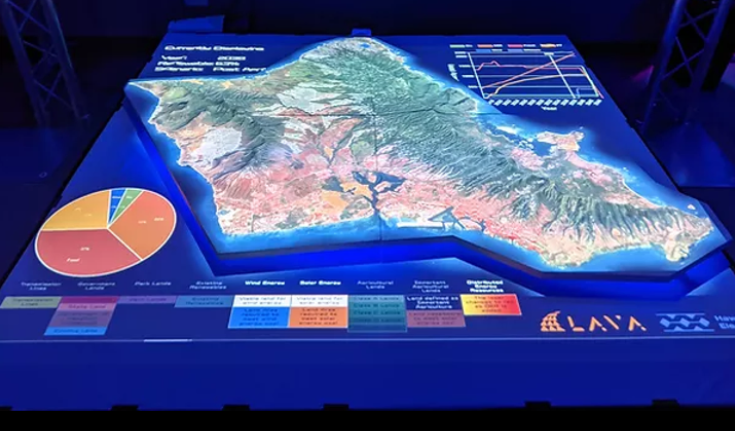}
    \includegraphics[width=\columnwidth]{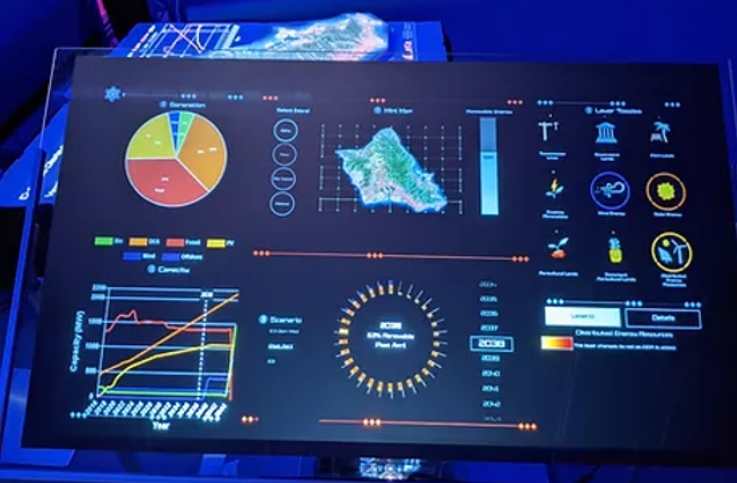}
 \caption{ (Top) The table display of the ProjecTable 2.0 system. (Bottom) The touch display of the ProjecTable 2.0 system with a very large touchscreen. }
 \label{fig:hecoTable}
\end{figure}

Building on the foundation of ProjecTable 1.0, this version retained the use of geospatial data layers over 3D topography but introduced additional functionality. Scenarios now included a wider range of energy data, such as solar, wind, and fossil fuel potentials, as well as identifying optimal deployment locations across Hawai‘i's islands. The system was designed to support updated and more dynamic scenarios, reflecting evolving strategies to achieve the renewable energy targets. ProjecTable 2.0 became a more comprehensive tool, expanding its use beyond public communication to include detailed planning and analysis tailored to local policymakers responsible for managing community development. See Figure~\ref{fig:hecoTable} for the full display of the ProjecTable 2.0 system.

\begin{figure}[tb]
 \centering 
    \includegraphics[width=\columnwidth]{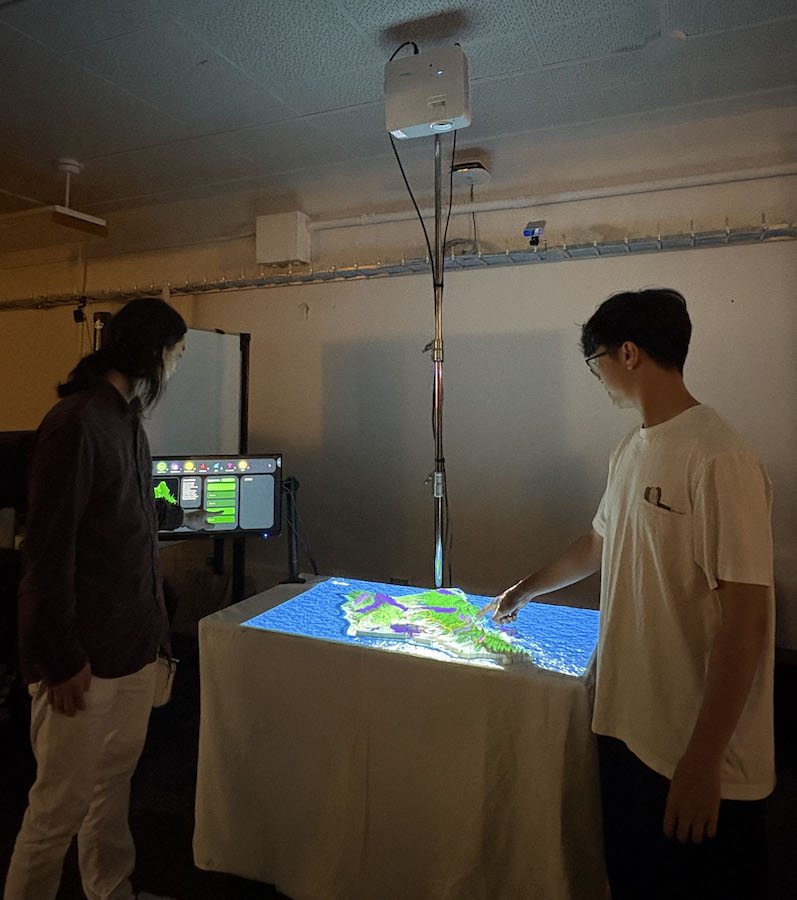}
        \includegraphics[width=\columnwidth]{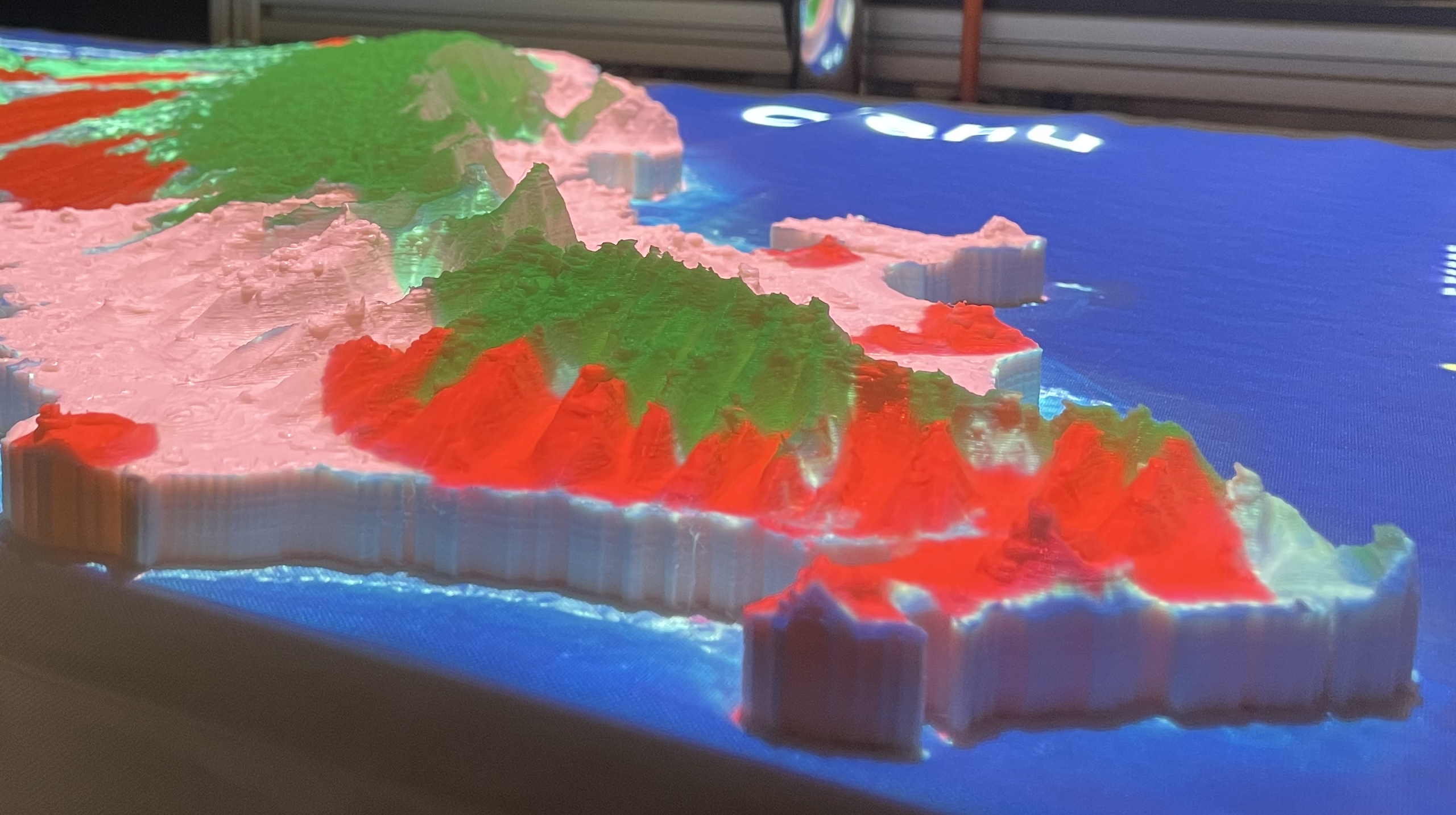}
 \caption{ (Top) The ProjecTable 3.0 system in action, showing a user interacting with the touchscreen display to toggle the Agriculture sub-layers, while another user observes the data. (Bottom) A close-up view of the 3D model with data layers projected onto it. }
 \label{fig:1}
\end{figure}

\subsection{Hardware}
\begin{figure}[tb]
 \centering
 \includegraphics[width=\columnwidth]{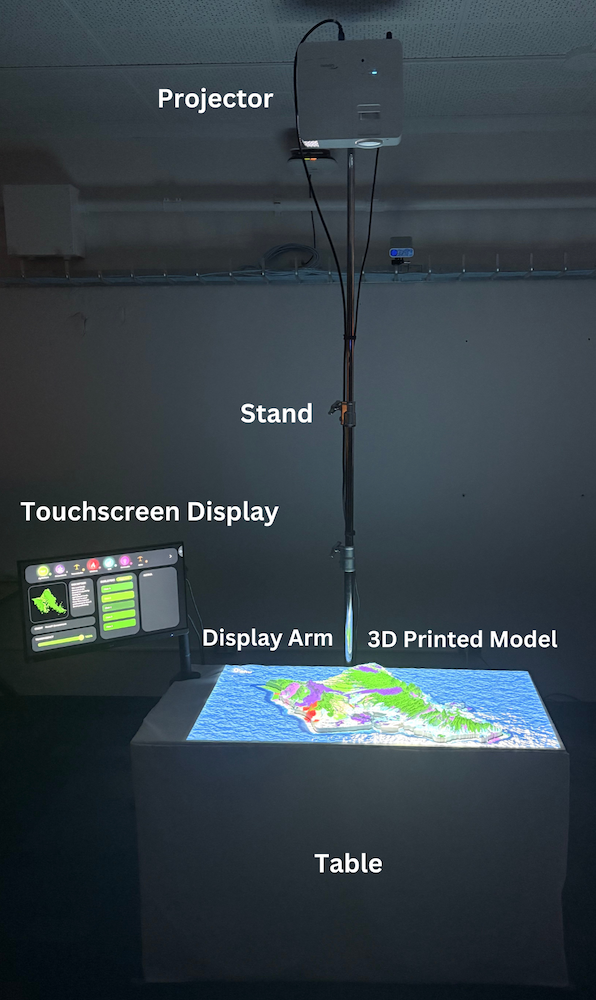}
 \caption{Hardware components of the ProjecTable 3.0 system, labeled.}
 \label{fig:system}
\end{figure}

The ProjecTable 3.0 setup includes a 4ft x 2ft foldable table, a touchscreen display mounted on an adjustable arm that is clamped to the table, a heavy-duty projector stand, a projector, and a laptop to run the software. The projector stand, typically used in the performance industry, can fold flat to the table's dimensions for easy transport. The material used to print the 3D models is PETG, a strong and durable material. To cover the table, we use a white sheet that improves projection quality. Excluding the 3D models, the system components can be purchased from retailers for around \$5000, depending on the quality of the laptop. The 3D models were printed in-lab using a Modix V3 BIG-120X printer, but similar prints can be outsourced or produced in parts using smaller printers. The complete ProjecTable setup is designed for portability and durability, ensuring it can be deployed in a variety of settings. See Figure~\ref{fig:system} for an illustration of the ProjecTable system and Figure~\ref{fig:1} for the ProjecTable system being used in-lab.

\subsection{Software}
The software was created using the Unity Game Engine. We chose Unity because it allows for easy transportability of the executable across different devices. This means that all the Makawalu VE tools can be zipped and opened on various devices without complications, making it possible for users to share content with each other. In this section, we describe the foundation of all the Makawalu VE tools from which they are derived. At its core, the software is organized within scenes, and content meant for 2D displays is added to a canvas. In ProjecTable, the scene uses two canvases: one for the touchscreen display and one for the projector display, which projects onto the table with the 3D model.

\section{Makawalu VE Tools Overview}
The word ``Makawalu'' in Hawaiian means numerous, in great quantities, or ``eight eyes'' \cite{makawaluMeaning}. It can also represent various other concepts, such as the need to view real-world problems and solutions through multiple lenses and angles simultaneously \cite{pukuiDictionary}. The main project lead for Makawalu, Kari Noe, decided to name the project and software "Makawalu" to reflect the software’s intended purpose: to provide the ProjecTable architecture with the ability to present and analyze many perspectives and forms of data in relation to a place.

\begin{figure*}[tb]
 \centering
    \includegraphics[width=0.6\textwidth]{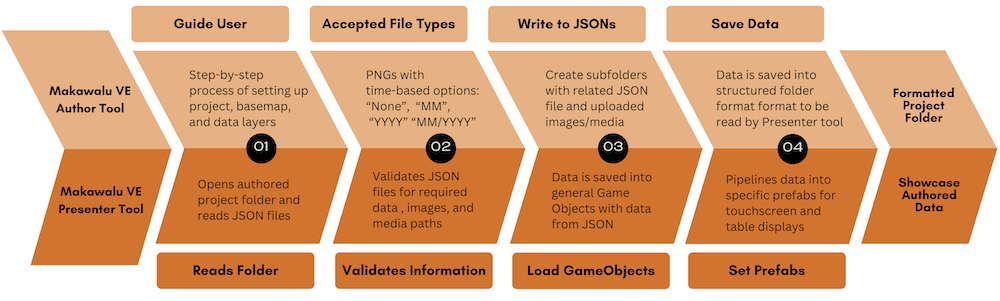}
 \caption{ A high-level diagram of the Makawalu VE Author and Presenter tool's architecture. }
 \label{fig:architecture}
\end{figure*}

The decision to separate Makawalu VE into three distinct tools allows each to function independently, catering to the specific needs of users. Together, these tools form a cohesive three-part system that enables Makawalu VE to evolve as a platform without requiring extensive developer involvement. The Author Tool enables users with minimal coding experience to input their data through a step-by-step process, creating their own authored version of ProjecTable. It generates a complete, formatted project folder containing their data. The Presenter Tool opens the authored project folder, reads the data, and pipelines it into delegated GameObjects based on the data type, allowing users to seamlessly showcase their data. The high-level architecture of both the Makawalu VE Author and Presenter Tools is shown in Figure~\ref{fig:architecture}, providing an overview of how these tools interconnect to form a cohesive system. Keeping the project folder as a separate component also makes it easy and quick for users to access the project on multiple computers. Lastly, the Editor Tool which will be briefly discussed, is designed for users with coding experience, allowing them to create custom components within the ProjecTable and implement and test them according to their needs. The features created through this tool could potentially be added to future updates of the Makawalu VE Tools.

\subsection{Makawalu VE Author Tool}
The Makawalu VE Author Tool is designed for users with little to no coding experience who wish to create their own version of ProjecTable. The tool provides an intuitive step-by-step workflow to guide users in entering data and uploading files directly into the application.

\subsubsection{Workflow and Functionality}
\begin{figure}[tb]
 \centering
    \includegraphics[width=\columnwidth]{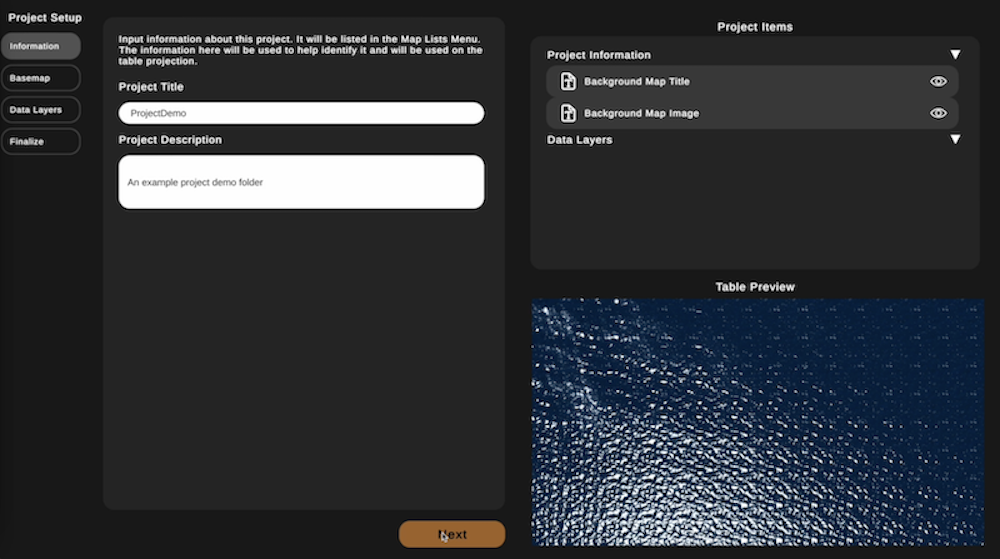}
 \caption{The first step when creating an authored ProjecTable. The project name and project description is filled out. }
 \label{fig:3}
\end{figure}

\begin{figure}[tb]
 \centering 
    \includegraphics[width=\columnwidth]{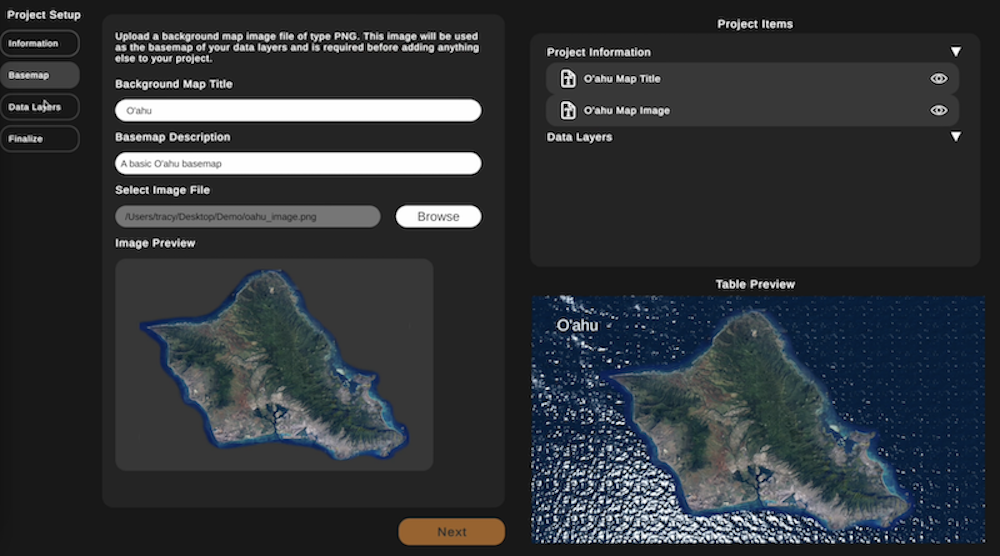}
 \caption{ The second step to creating an authored ProjecTable is the basemap setup. Basemap information entered is about O`ahu. A basemap png image of O`ahu is uploaded. }
 \label{fig:4}
\end{figure}

\begin{figure}[tb]
 \centering 
    \includegraphics[width=\columnwidth]{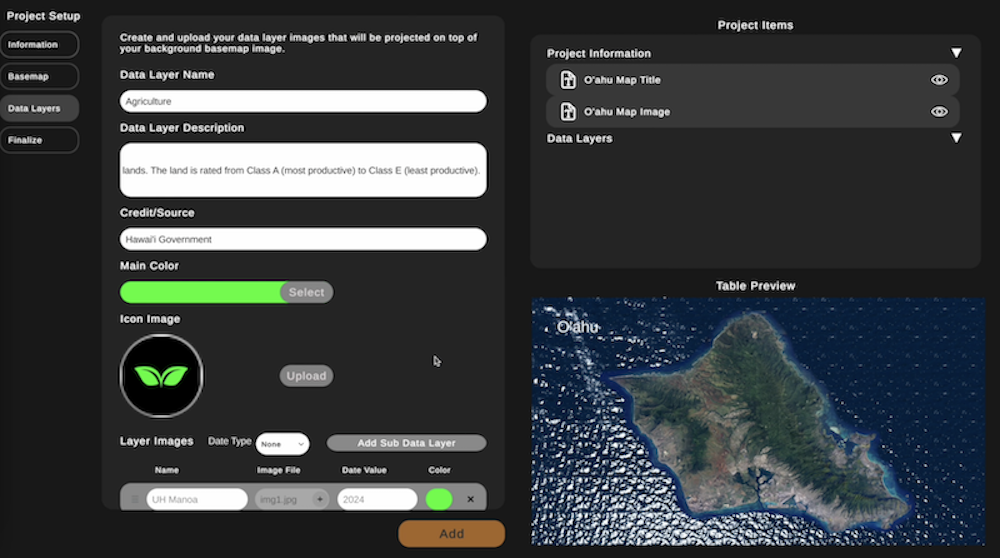}
    \includegraphics[width=\columnwidth]{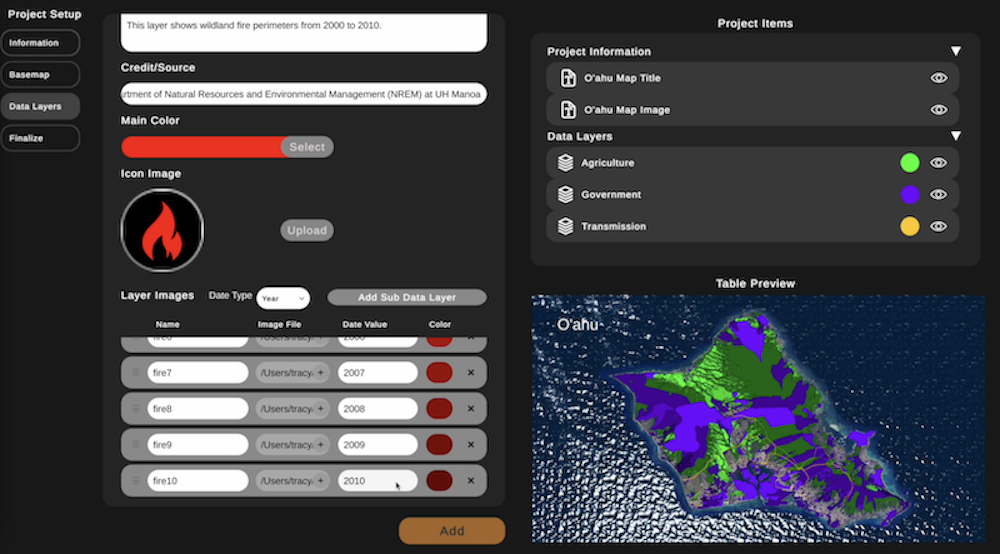}
 \caption{(Top) The third step in creating an authored ProjecTable which is the data layer step. Agriculture information is filled in and the data layer color and icon are also selected. (Bottom) A wildfire data layer is being created with time-based sub layers where the date type is by years. }
 \label{fig:5}
\end{figure}

\begin{figure}[tb]
 \centering
    \includegraphics[width=\columnwidth]{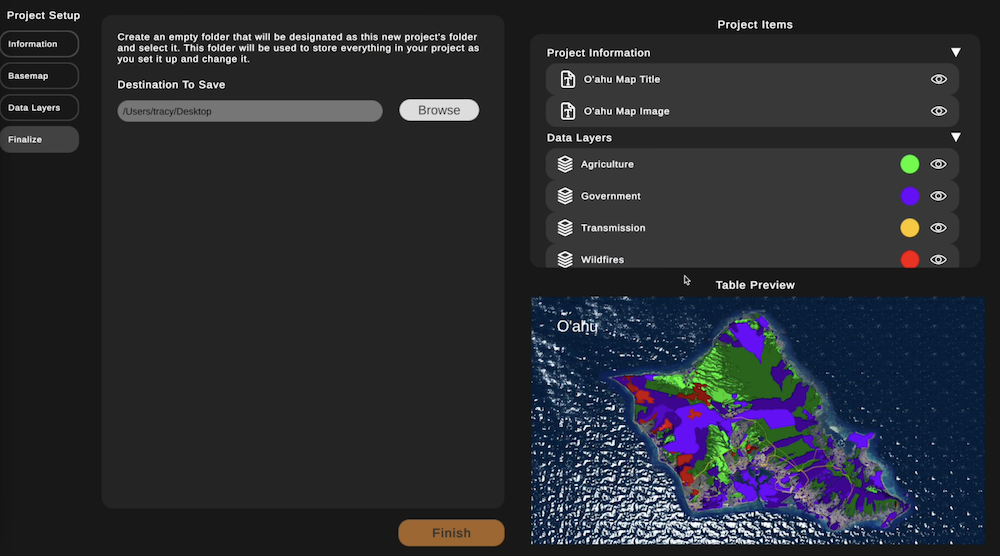}
 \caption{ The last step which prompts the user to select a destination on the computer of where to save the project folder. }
 \label{fig:6}
\end{figure}

The process begins with naming the project, followed by adding basemap information and uploading a corresponding image. Users then configure multiple data layers by providing details such as the layer name, description, credit or source, an icon to represent the data layer, and a color. The tool supports time-based formats for data layers, including month, year, or month/year, as well as static data without time associations. This functionality allows users to iterate through time-based data seamlessly, rather than manually toggling individual layers. Once all data is entered, users finalize their project by selecting a destination on their computer to save the project folder. This process is illustrated in Figures~\ref{fig:3}, \ref{fig:4}, \ref{fig:5}, and \ref{fig:6}, which depict the steps to name the project, add information from the basemap, configure the data layers and finalize the project, respectively.

When the project is saved, the Author Tool organizes all inputted information into a structured folder system. Subfolders are created to store images and other assets, while all user-provided information is saved in JSON files. These files include paths to the associated data layer images, ensuring proper organization and functionality. The JSON files are also structured to be compatible with the Presenter Tool, making the project ready for immediate use. The entire project folder can be compressed into a zip file and shared across platforms for easy deployment.

\subsubsection{User Interface Design}
The user interface of the Author Tool is designed for accessibility on a standard computer, although it is most effective when used with the appropriate hardware and software components of the ProjecTable. The interface features several panels to facilitate project creation.

A navigation list on the left-hand side outlines the steps, allowing users to input information systematically. The adjacent right-hand panel provides a real-time preview of the project, which updates automatically as the user progresses through each step. An upper panel displays a summary of the project information, including the project name, basemap name, and data layer details such as names, colors, and sublayers. At least one sublayer is required for each data layer. Additionally, a bottom panel offers a mini table view of the project display, which updates dynamically as users make changes.

\subsubsection{Flexibility and Real-Time Editing}
The Author Tool allows users to revise any part of their input during the creation process. Changes are immediately reflected in the real-time preview and summary panels. This flexibility ensures that the project remains consistent with the user's vision, while maintaining an organized and user-friendly workflow.

\subsection{Makawalu VE Presenter Tool}
The Makawalu VE Presenter Tool is the main application to display data from a project folder, enabling the ProjecTable system to display custom visualizations rather than hard-coded data. Upon launch, the tool prompts the user to select a project folder and reads the JSON files within it. During this process, the tool validates the presence of all necessary files and ensures that all file paths are valid, with images properly stored in their respective folders. The user interface (UI) is built using Unity assets, such as the Modern UI Pack \cite{unityModernUI}, to enhance interaction and presentation.

Once the validation process is complete, the tool loads the data into GameObjects, which contain the information and images read from it's respective folder. These GameObjects are then processed into prefabs based on the type of data they represent. The prefabs were built using the Modern UI Pack asset and tailored to meet the needs of Makawalu VE. Each data layer has its own separate window of prefabs containing related information that can be toggled through.

For example, data layers without time-based information display a window with default information and individual toggle buttons for activating or deactivating layers. In contrast, time-based data layers feature a window with default information and a scroll view, allowing users to navigate chronologically through the data. The Modern UI elements and scroll view components were developed using the Unity asset MzTools \cite{unityMzTools} and further customized to meet the specific requirements of Makawalu VE. This system was originally developed by a graduate student, Marissa Halim, who was also working in parallel on a different component of Makawalu VE. It was later implemented into the Makawalu VE Presenter Tool.

\subsubsection{Touchscreen Display}
\begin{figure}[tb]
 \centering 
    \includegraphics[width=\columnwidth]{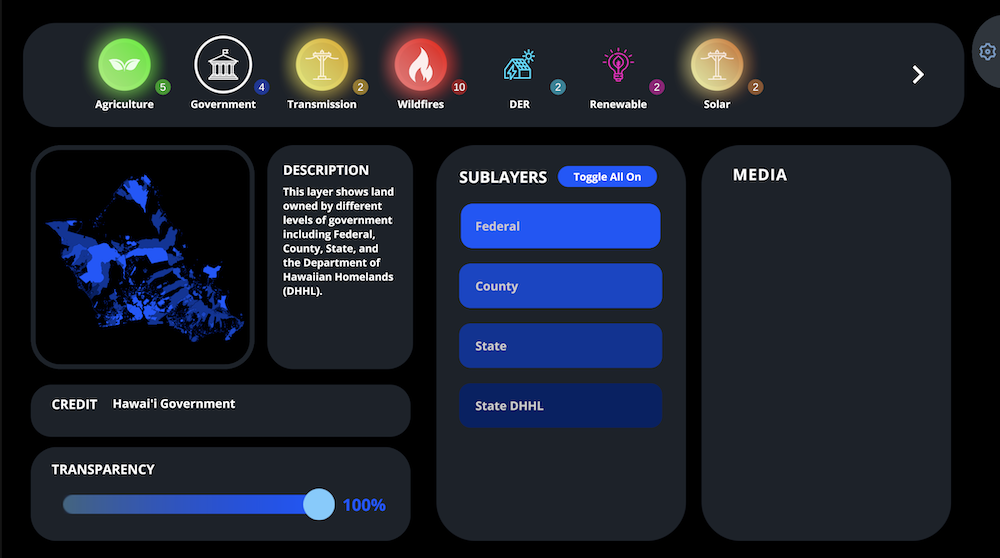}
 \caption{ The touchscreen display of the Makawalu VE Presenter Tool. The government data layer is displayed with sub layer buttons that are not time-based and can be individually toggled on and off. }
 \label{fig:7}
\end{figure}

\begin{figure}[tb]
 \centering 
    \includegraphics[width=\columnwidth]{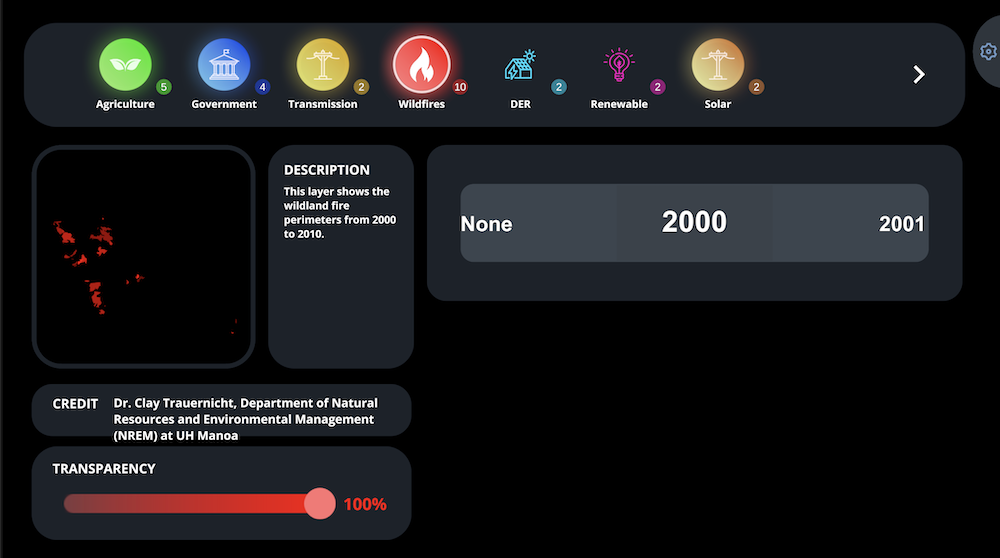}
 \caption{ The wildfire data layer is displayed with a time-based scrollview containing year data. The year 2000 is currently selected which will activate the matching data layer image onto the table display. }
 \label{fig:8}
\end{figure}

\begin{figure}[tb]
 \centering
    \includegraphics[width=\columnwidth]{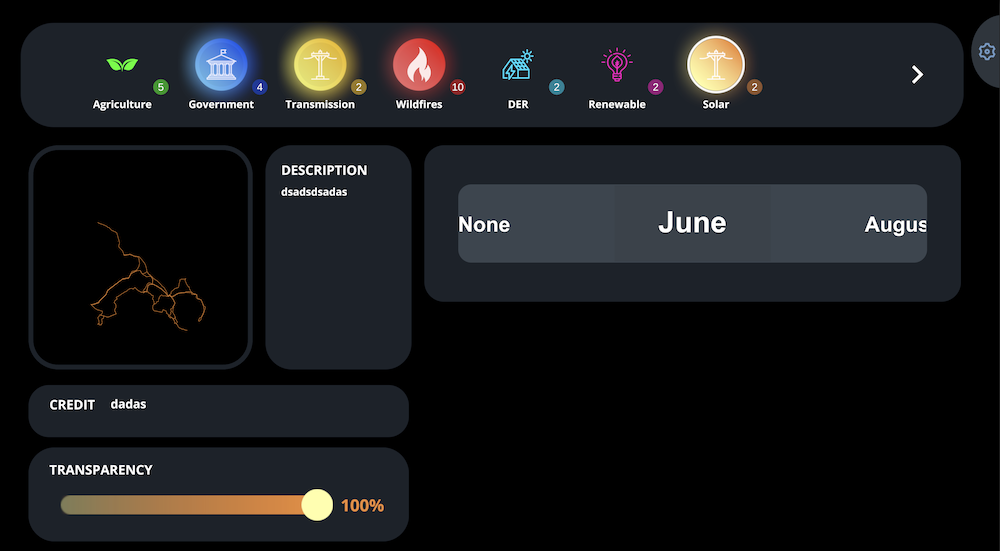}
 \caption{ The solar data layer information is displayed with a time-based scrollview containing month data. The month June is selected and will activate the corresponding data layer image into the table display. }
 \label{fig:9}
\end{figure}

\begin{figure}[tb]
 \centering
    \includegraphics[width=\columnwidth]{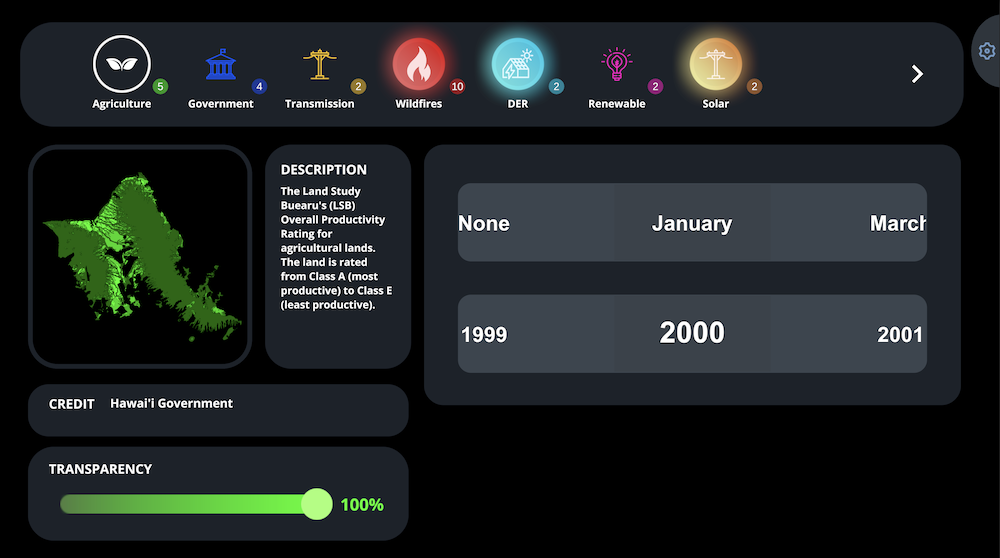}
 \caption{ The agriculture data layer information is displayed with a time-based scrollview containing year with month data. The year selected is 2000, and the month selected from the year 2000 is Janurary. The data layer corresponding to agriculture's data layer image for the date values 2000 and Janurary will activate on the table display. }
 \label{fig:10}
\end{figure}

The touchscreen display serves as the primary user interface for interacting with the table display. Upon initialization, the canvas defaults to a layout divided into three main sections:

\begin{enumerate}
    \item Top Section: This section contains toggle buttons for each data layer. Each button corresponds to a geospatial data layer that includes an icon, a legend, a description, and source/credit information. Clicking a toggle button updates the bottom section to display information related to the selected data layer.
    \item Bottom Section (Left-Side Panel): This panel displays detailed information about the currently toggled data layer, including: 
        \begin{enumerate}
            \item A small image preview of the data layer.
            \item A description of the data layer.
            \item The source/credit for the origin of the data.
            \item A transparency slider to adjust the opacity of the data layer on the table display.
        \end{enumerate}
    \item Bottom Section (Right-Side Panel): This panel varies based on the time-based type of the data layer which can include None, Month, Year, or Year with Months:
        \begin{enumerate}
            \item None: Displays individual sub layer buttons that allow users to toggle and layer any or all sublayers on or off (See Figure \ref{fig:7}).
            \item Time-Based Data Layers: Provides scroll views for navigating chronological data. For example:
                \begin{enumerate}
                    \item Month-Based Layers: Presents a scroll view with all the months (See Figure \ref{fig:9}).
                    \item Year-Based Layers: Presents a scroll view with all the years (See Figure \ref{fig:8}).
                    \item Year-Month-Based Layers: Includes two scroll views: the lower scroll view lists all years, and selecting a year updates the upper scroll view to show the corresponding months. Users can then scroll through the months, updating the table display to show the relevant data layer image that matches the selected year-to-month value (See Figure \ref{fig:10}).
                \end{enumerate}
        \end{enumerate}
    \item Media Panel: This placeholder panel will support additional media in future iterations, such as videos, audio, and related images associated with data layers.
\end{enumerate}

\subsubsection{Table Display} 
The table display is designed to augment the 3D-printed topography by presenting additional contextual information, graphs, and visuals to enhance the narrative the user aims to convey. Both the base layer and data layers, which overlay the topography, are created in QGIS, an open-source geographic information system (GIS) software. Geospatial datasets from HSEO were imported into QGIS and exported as PNG images for the base map and data layers using a Python script.

A satellite image is used as the base map, as its realism helps users connect the data to actual locations. To enhance the visual appeal, an animated water effect surrounds the island, creating an immersive experience for viewers.

All table display components, including the base map and overlaid data layers, are adjustable via a settings page accessible on the touch display. Exhibition operators can move and resize elements to customize the presentation. However, it is recommended that this feature not be used once a demonstration has begun to avoid disrupting the display.

\section{Evaluation}
The latest ProjecTable system has been presented at numerous events this year which has provided a lot valuable feedback. These events include the NASEO and NGA events that were both held at LAVA. Additionally, HSEO has taken the ProjecTable to various outreach events such going to Moloka`i to showcase data that is relevant to local energy needs. Feedback was gathered from this even to improve the system's usability to ensure it effectively meets the needs from different communities. LAVA has also taken the ProjecTable to events such as the 13th Festival of Pacific Arts and Culture 2024 and A Catalyst for Change: Indigenous Innovation Conference 2024. 

\begin{table*}[ht]
\centering
\begin{tabular}{ |c|c|c|c|c| }
 \hline
 \multicolumn{5}{|c|}{Survey Results} \\
 \hline
 Survey Question & Not at all (1) & Very little (2) & Somewhat (3) & To a great extent (4)\\
 \hline
 Improved your interest in the data & 0 & 0 & 0 & 10\\
 Improved your understanding of the data & 0 & 0 & 0 & 10\\
 Helped you answer questions about the data & 0 & 0 & 1 & 9 \\
 Helped you come up with new insights & 0 & 0 & 1 & 9\\
 Ease of remembering data with HAVENTable & 0 & 0 & 1 & 9\\
 Interest in future use of HAVENTable & 0 & 0 & 1 & 9\\
 \hline
\end{tabular}
\caption{Survey results on the effectiveness of the HAVENTable, focusing on user interest, understanding, and potential future use.}
\label{tab:survey-results}
\end{table*}

\subsection{Survey Feedback}
For the Moloka`i outreach event, HSEO conducted a survey to gather community opinions regarding energy sustainability efforts on Moloka`i. The survey aimed to assess the public's awareness of renewable energy options and their willingness to support different energy strategies. It was distributed to local residents and policymakers through both online platforms and in-person community meetings. The questions focused on topics such as energy preferences, attitudes toward renewable resources, and perceptions of the potential environmental impacts of energy development. The survey consisted of multiple-choice questions, Likert scale ratings, and open-ended responses to allow for both quantitative and qualitative insights (See Table~\ref{tab:survey-results} for survey results). They were able to gather survey results from 10 participants. On average, each question received a rating of 9 out of 10, indicating positive experiences. Based on the results of the survey, the latest ProjecTable system received very positive feedback for its effectiveness in helping users understand the data and gain new insights into their local community's energy and available resources.

\subsection{Survey Open-Ended Questions}
For the open ended questions, they were asked "Can you give an example of a question you had that was answered or an insight you gained because of the large 3D island model?". The survey responses highlighted key insights from the use of the large 3D printed model. Participants noted the model effectively answered questions related to land ownership, environmental impact and project practicality. They mentioned they gained clarity on property ownership, such as identifying small parcels owned by the Moloka`i Ranch. Others responded that it helped them understand the interplay between land ownership, grid hookups, and wildfire risks. The ProjecTable also helped participants evaluate areas with more moisture and better understanding of the topography of the island. The realistic satellite imagery helped provide a more tangible perspective on how proposed projects could impact Moloka`i's natural environment, resources, and cost.

\subsection{Survey Suggestions and Improvements}
The feedback for improvements on the system highlighted strong interest in expanding and enhancing the system's capabilities. Suggestions included adding more data layers, in particular, for energy, water, sea-level rise, to support various scenarios. Participants also expressed a desire for larger displays, potentially wall-mounted version, and increase the number of tables that are available for use. One response emphasized spending time with knowledgeable guides who know about the table's information to maximize the table's effectiveness. A more specific request was made for the Moloka`i community to have its own table which shows enthusiasm for having more tables for broader accessibility among their local communities. Feedback for suggestions was insightful and overwhelmingly positive in experience with participants expressing excitement about its potential use in the future in representing their own community.  

\section{Future Work}
\subsection{Standardized UI Components}
The Makawalu VE tools are still in the early stages of development, with the main goal being to create a more modular version of the ProjecTable. We have implemented a few UI components, as seen in the Makawalu VE Presenter Tool, to display general information about a data layer. However, we acknowledge that it does not yet fully cover all the necessary aspects to be completely modular.

As we continue to gather feedback from users with different data sets and needs, we will assess the most common use cases. This will enable us to develop more modular components within the Makawalu VE tools to better accommodate users with similar types of data. For example, the scroll view is a newly standardized UI component created to support high volumes of time-based data. It offers smoother interaction than a traditional UI slider, making it ideal for enabling users to navigate through time-based data in a more intuitive and seamless way.

\subsection{Story Maps}
Throughout the development of the Makawalu VE tools, there have been numerous questions about whether the ProjecTable or Presenter Tool has a way to story-map data. This feature would allow users to navigate through an authored narrative instead of toggling individual data layers on and off to interpret the data. It could potentially become another standardized UI component to add to the Makawalu VE tools, especially for users who prefer a PowerPoint-style display.

\subsection{Merging LAVA Students' Contributions}
Research assistants from LAVA have been working in parallel to create various ProjecTable projects for clients interested in the tool. These clients include the Hawai'i Climate Data Portal (HCDP), focusing on displaying rainfall data; National Aeronautics and Space Administration (NASA), aiming to showcase sea-level rise data at the Kennedy Space Center; and Hawai'i Climate Smart Commodities (HiCSC), which seeks to help farmers optimize methods based on land type through data visualization.

Due to the high demand from clients and the ongoing development of the Makawalu VE tools, LAVA research assistants have had to split their efforts to address the needs of users with different datasets. This process enabled the development of new ProjecTable components, which will be later integrated into Makawalu VE tools. All these new features and components will ultimately contribute to improving the functionality of the Makawalu VE tools.

\subsection{AI-Driven Layout in the Presenter Tool}
The Makawalu VE Presenter Tool features a touchscreen display designed with modularity in mind, allowing developers to modify the shape, position, and size of its components. These components refer to any objects contained within the rectangular or square boxes on the touchscreen display. For example, the display shown in figure~\ref{fig:10} would have a total of six components. These components can be customized according to the developer's preferences for what to include in the display. Currently, the tool has a default layout with three main sections to display data layer information. This layout is adjustable, enabling users to reposition or remove components as needed. Additionally, we could develop several general layouts based on the types of data being displayed, such as with the scroll view, to meet the preferences of most users.

The Presenter Tool could also incorporate AI to optimize the layout orientation according to the user-input data layer. This would reduce the need for developers to create new layouts each time. For example, when the tool reads the project folder for the first time, it generates a display that it believes best fits the inputted data layer. If the user isn’t satisfied, they can easily revert to the default layout or select from a list of AI-generated layouts tailored specifically to that data layer.

\section{Conclusion}
In settings where the Presenter Tool is used in public spaces, such as schools or conferences, we aim to facilitate the easy sharing of ProjecTable versions. Users would only need their own ProjecTable system set up, the Presenter Tool downloaded, and the project folder to load and display their own copy of an authored ProjecTable version from someone else. This approach promotes seamless sharing and accessibility, ensuring that the Makawalu VE system can be easily replicated and utilized across various platforms. By integrating the flexibility of the Makawalu VE tools with the simplicity of sharing and customization, we empower users to create and communicate data-driven stories more effectively. As the system evolves, we continue to prioritize modularity and user accessibility, allowing for greater adoption and adaptation in diverse environments.

\acknowledgments{
This work was enabled in part by funding from the National Science Foundation awards: 2139133, 2201428, 2232862, 2004014, 2003800, 2003387, 2117975, 2138259, 2138286, 2138307, 2137603, and 2138296, as well as support from the Academy for Creative Media System. We would like to acknowledge the Hawai‘i State Energy Office (HSEO) for their continued collaboration.}

\bibliographystyle{abbrv-doi}

\bibliography{template}
\end{document}